%% file: NCCQFT2.tex
\documentclass[a4paper,
				headsepline,
				headinclude=false,
				footinclude=false,
				headsepline,
				fontsize=12pt]{scrartcl}

\input{Pakete.tex}

\usepackage[automark]{scrpage2}

\numberwithin{equation}{section}	
\graphicspath{{img/}}

\allowdisplaybreaks
\begin{document}
	{\centering	\Large
\textbf{Noncommutative 3-colour scalar quantum field theory model in 2D}		
\par
	\vspace*{4ex}
\large{Alexander Hock, Raimar Wulkenhaar}\\
\vspace*{4ex}
\small
 \textit{Mathematisches
Institut der Westfälischen
Wilhelms-Universität,\\
Einsteinstraße 62, D-48149 Münster,Germany}\\
\vspace*{2ex}
\small{E-mails: a\_hock03@uni-muenster.de, raimar@math.uni-muenster.de}}\\
\vspace*{4ex}
\hrule \vspace*{2ex}
\textbf{Abstract}

\vspace*{2ex} We introduce the 3-colour noncommutative quantum field
theory model in two dimensions. For this model we prove a generalised 
Ward-Takahashi identity, which is special to coloured noncommutative
QFT models. It reduces to 
the usual Ward-Takahashi identity in a
particular case. The Ward-Takahashi identity is used to simplify the
Schwinger-Dyson equations for the 2-point function and the $N$-point
function. The absence of any renormalisation conditions in the 
large $(\mathcal{N},V)$-limit in 2D leads
to a recursive integral equation for the 2-point function, which we 
solve perturbatively to sixth order in the coupling constant.

\vspace*{2ex}
\hrule
\vspace*{2ex}

\pagenumbering{arabic}
\section{Introduction} \setcounter{page}{1} Consider a hexagonal
lattice with three different coloured links, where at each vertex all
three links carry different colours. The mathematical problem of
counting the number of colourings of a lattice with $N$ vertices was
solved by Baxter \cite{Baxter1970}.  A generalisation of this
so-called 3-colour model as a Hermitian matrix model problem was
introduced by Eynard and Kristjansen \cite{Eynard1998} and solved by
Kostov \cite{Kostov2002}.  Eynard and Kristjansen reduced the 
partition function (without external fields) to an integral over
eigenvalues, which could be solved by saddle-point techniques.

Graphs in $\mathcal{N}{\times}\mathcal{N}$-matrix models are ribbon
graphs on a Riemann surface.  These ribbon graphs are dual to the
triangulations of the corresponding surface.  The large-$\mathcal{N}$
limit is dominated by planar graphs, corresponding to triangulations
of the sphere.  Two-dimensional quantum gravity can be formulated as a
counting problem for triangulations of random surfaces, which leads to
the connection between 2D quantum gravity and random matrices
\cite{Douglas1990,Gross1990}.

Moreover, it was proved by Kontsevich \cite{Kontsevich1992} that the
solution of an action of the form $\mathrm{tr}(E\cdot
  M^2+\frac{\mathrm{i}}{6}M^3)$ with Hermitian matrices $M$ and
external matrix $E$ can be mapped to a Hermitian matrix model with arbitrary
potential. On the other hand the Kontsevich model proves Witten's 
conjecture about intersection numbers of stable cohomology classes on 
the moduli space of curves \cite{Witten1990}. 
One particularly elegant solution technique reduced the partition
function to an integral over the eigenvalues $x_i$ and observed 
that these integrals are unchanged under diffeomorphisms of $x_i$
generated by $x_i^{n+1}\frac{d}{dx_i}$. The corresponding Virasoro
constraints all descend from a master constraint which was solved by
Makeenko-Semenoff \cite{MAKEENKO1991}.

Matrix models gained renewed interest in a non-perturbative approach
to quantum field theories on Moyal-Weyl deformed noncommutative space
\cite{Langmann:2003if,Grosse:2005ig}. These approaches use the matrix
basis of the Moyal space and add a harmonic oscillator potential to
the Laplacian \cite{Grosse2005a}. The most established noncommutative
quantum field theory is the $\Phi^4$-model \cite{Grossea}, which is a
candidate for an exactly solvable quantum field theory in 4D due to
its vanishing $\beta$-function at all orders \cite{Disertori2007}.
Recently all boundary sectors of  
noncommutative $\Phi^3$-model in $\{2,4,6\}$ dimensions 
were solved exactly in the large $(\mathcal{N},V)$-limit 
\cite{Grosse2017,Grosse2018}.

In this paper we will study the \textit{noncommutative 3-colour model}
as a quantum field theoretical model.  Roughly speaking, it is the
model solved by Kostov with an additional external dynamical field $E$
of linearly-spaced eigenvalues.  Although it shares topologically some graphs
with the noncommutative $\Phi^3$-model \cite{Grosse2017}, it has more
similarities to the $\Phi^4$-model \cite{Grossea} due to the absent
1-point function.  For the large $(\mathcal{N},V)$-limit a
closed integral equation for the 2-point function will be derived. In
the two-dimensional case the first-order loop correction
has no UV divergence so that 
this 2D noncommutative 3-colour model needs no
renormalisation in this limit. The closed integral equation defines
a recursion formula for its perturbative expansion. 
Absence of any renormalisation makes this recursion easy. We
are able to determine perturbatively the 2-point function up to the 
sixth order in the coupling constant.  

The action of the noncommutative 3-colour model for real scalar fields
$\phi^a$ with colour $a\in\{1,2,3\}$ is given by
\begin{align}\label{action}
 S[\phi]:=\frac{1}{8 \pi}\!\int_{\mathbb{R}^2} \!\!dx \!\left(\frac{1}{2}\sum_{a=1}^3\phi^a\!\left(-\Delta +\|2\Theta^{-1}x\|^2+\mu^2\right)
 \phi^a\!+\!\frac{\lambda'}{3}
 \!\!\sum_{a,b,c=1}^3\!\!
\sigma_{abc} \phi^a\star \phi^b\star\phi^c\!\right)\!(x),
\end{align}
where $\sigma_{abc}=1$ for $a\neq b\neq c\neq a$ and $\sigma_{abc}=0$ else.
Here $\lambda'\in\mathbb{R}$ is the coupling constant, $\mu^2$ the 
mass squared and $\Delta$ the Laplacian, where
independence of any colour is assumed. 
The Laplacian has non-compact resolvent,
therefore the harmonic oscillator potential is added 
to achieve compactness. The Moyal $\star$-product is defined by
\begin{align*}
 (f\star g)(x)=\int_{\mathbb{R}^2\times\mathbb{R}^2} \frac{dy \,dk}{(2\pi)^2}f\left(x+\tfrac{1}{2}\Theta k\right) g\big(x+y\big)
 e^{\mathrm{i}\langle k,y\rangle},\qquad f,g\in \mathcal{S}(\mathbb{R}^2),
\end{align*}
where $\Theta$ is a $2\times 2$ skew-symmetric matrix with
$\Theta_{12}=-\Theta_{21}=:4 V>0$.

The vector space $\mathcal{S}(\mathbb{R}^2)$ equipped with the Moyal
$\star$-product has a matrix basis $f_{nm}(x)$, which is the
2-dimensional harmonic oscillator basis (independent of the 
harmonic oscillator term in (\ref{action}), although 
equality of both frequencies reduces a tensor product 
to a matrix product). The
formulation of the action in the matrix basis is obtained from the
expansion
\begin{align*}
 \phi^a(x)=\sum_{n,m=0}^\infty \phi^a_{nm}f_{nm}(x),\qquad\text{with}\qquad x\in \mathbb{R}^2.
\end{align*}
The matrix basis satisfies \cite{Grosse2003}
\begin{align*}
 (f_{nm}\star f_{kl})(x)=\delta_{mk}f_{nl}(x),\qquad \int_{\mathbb{R}^2}dx f_{nm}(x)=8\pi V\delta_{nm}.
\end{align*}
Accordingly, the action in the matrix basis with an UV cut-off $\mathcal{N}$ is given by
\begin{align}\label{actionmatrix}
 S[\phi]&=V\left(\sum_{a=1}^3\sum_{n,m=0}^\mathcal{N}\frac{H_{nm}}{2}\phi^a_{nm}\phi^a_{mn}+\frac{\lambda'}{3}
		\sum_{a,b,c=1}^3\sum_{n,m,l=0}^\mathcal{N}\sigma_{abc}\phi^a_{nm}
		\phi^b_{ml}\phi^c_{ln}\right)\\
		H_{nm}&:=E_n+E_m,\qquad E_m=\frac{m}{V}+\frac{\mu^2}{2},\nonumber
\end{align}
where $(\phi^a_{nm})$ are Hermitian matrices. The linear and discrete
dependence of $E_m$ reflects the eigenvalue spectrum of the
quantum-mechanical harmonic oscillator.

\section{Graph computation}\label{SectionGraph}
As a perturbative theory the noncommutative 3-colour
model can be expanded by graphs. In the large $(\mathcal{N},V)$-limit 
taken later, all functions of genus $>0$
are suppressed, which is a well-known behaviour for matrix models.
Therefore we will consider only planar ribbon graphs, but we admit 
a non-trivial boundary structure for which we next give an alternative
description. 

Let $\Gamma$ be
a planar ribbon graph on $S^2$ consisting of vertices, edges and faces
subject to the following conditions: 
It has two different vertices, black
(internal) and white (external) vertices. The external vertex is also
called boundary component. The number of white vertices is $B\geq 1$,
and each white vertex has the valence $N_\beta$ for $\beta\in
\{1,...,B\}$. The edges have one of three different colours; they
separate two faces. The black vertices are of valence three. At
a black vertex all three colours must occur once. We require that every
face has at most one external vertex, in which case it is called
external. If the $\beta^{\mathrm{th}}$ external vertex has valence
$N_\beta$ it is a corner of $N_\beta$ external faces which are 
labelled by positive numbers $p_1^\beta,...,p_{N_\beta}^\beta$. Let 
$a_i^\beta$ be the colour of the edge which ends at the 
$\beta^{\mathrm{th}}$ external vertex and separates the faces 
labelled by $p_i^\beta$ and $p_{i+1}^\beta$, where
$i\in \{1,...,N_\beta\}$ and $N_\beta+1\equiv 1$.
Faces without an external vertex are called internal and are 
labelled by $q_1,...,q_L$.

To every white vertex a weight $1$ is associated and to every black
vertex a weight $\lambda$. An edge is weighted by
$\frac{1}{1+z_1+z_2}$ if $z_1$ and $z_2$ are the labels 
of the two faces separated by the edge. 

To a graph $\Gamma$ we associate the function
$\tilde{G}^{a_1^1...a^1_{N_1}|...|a_1^B...a^B_{N_B}}_{p_1^1...p^1_{N_1}|...|p_1^B...p^B_{N_B}}(\Gamma)$
given by multiplication of all weights of vertices and faces of
$\Gamma$ and with integration over the labels $q_1,...,q_L$ of
all internal faces from 0 to $\infty$.  We consider two graphs $\Gamma,\Gamma'$ as
equivalent, $\Gamma\sim \Gamma'$, if they are topologically the same
and have the same labels $
p_1^1,...,p^1_{N_1},...,p_1^B,...,p^B_{N_B}$ and
$a_1^1,...,a_{N_1}^1,...,a_1^B,...,a^B_{N_B} $, but different
assignment of colours at internal edges. Such graphs $\Gamma\sim
\Gamma'$ have the same amplitude
$\tilde{G}^{a_1^1...a^1_{N_1}|...|a_1^B...a^B_{N_B}}_{p_1^1...p^1_{N_1}|...|p_1^B...p^B_{N_B}}
(\Gamma)=\tilde{G}^{a_1^1...a^1_{N_1}|...|a_1^B...a^B_{N_B}}_{p_1^1...p^1_{N_1}|...|p_1^B...p^B_{N_B}}(\Gamma')$. We
denote by $s(\Gamma):=|[\Gamma ] |$ the number of graphs equivalent to
$\Gamma$.

For a fixed number $B$ of external vertices of valences $N_1,...,N_B$,
but arbitrary number of internal vertices, one can ask whether the sum over
all possible planar graphs converges for sufficient small
$|\lambda|$. This sum can formally be defined over all equivalence
classes by 
\begin{align*}
 G^{a_1^1...a^1_{N_1}|...|a_1^B...a^B_{N_B}}_{p_1^1...p^1_{N_1}|...|p_1^B...p^B_{N_B}}:
	=&\sum_{[\Gamma]\in \mathcal{G}^{\mathbf{a}}_{\mathbf{p}}}s(\Gamma)\tilde{G}^{a_1^1...a^1_{N_1}|...|a_1^B...a^B_{N_B}}_{p_1^1...p^1_{N_1}|...|p_1^B...p^B_{N_B}}(\Gamma),
\end{align*}
where $\mathcal{G}^{\mathbf{a}}_\mathbf{p}$ is the set of equivalence classes
of all planar graphs with $B$ external vertices, external edges of colour $a_1^\beta,...,a_{N_\beta}^\beta$ and 
external faces labelled by $p_1^\beta,...,p^\beta_{N_\beta}$ for all $ \beta\in \{1,...,B\}$. This sum can clearly be rearranged as a series in $\lambda$
\begin{align}\label{Gloopexpansion}
 G^{a_1^1...a^1_{N_1}|...|a_1^B...a^B_{N_B}}_{p_1^1...p^1_{N_1}|...|p_1^B...p^B_{N_B}}=:
 \sum_{n=0}^\infty \lambda^n G^{\quad\!\! a_1^1...a^1_{N_1}|...|a_1^B...a^B_{N_B}}_{n,\,p_1^1...p^1_{N_1}|...|p_1^B...p^B_{N_B}}.
\end{align}

\addsec{\large{Examples}}

$\tilde{G}^{a_1^1|a_1^2}_{p_1^1|p_1^2}\,\,\bigg(\qquad \quad \qquad \quad \bigg)$
\vspace*{-6ex}

\hspace*{5ex}
\begin{tikzpicture}
\draw [draw=red] (3.1,0) .. controls (3.6,0) and (3.1,0).. (3.6,0) node at (3.3,0.3) {\tiny $p^1_1$};
\draw [draw=red] (4.6,0) .. controls (5.1,0) and (4.6,0).. (5.1,0) node at (4.2,0.2) {\tiny $p^2_1$};
\draw [draw=blue] (3.7,0.1) .. controls (4,0.6) and (4.9,0.6).. (5.2,0.1);
\draw [draw=green] (3.7,-0.1) .. controls (4,-0.6) and (4.9,-0.6).. (5.2,-0.1);
\draw (3,0) circle (0.1cm);
\draw (4.5,0) circle (0.1cm);
\filldraw (3.7,0) circle (0.1cm);
\filldraw (5.2,0) circle (0.1cm);
\end{tikzpicture}
\vspace*{-8.5ex}
\begin{align*}
	\!\qquad\qquad=\frac{\lambda^2}{(1+p^1_1+p^2_1)^2(1+2p^1_1)(1+2p^2_1)}
\end{align*}
\vspace*{4ex}
$\tilde{G}^{a_1^1a_2^1}_{p_1^1p_2^1}\,\,\bigg(\qquad \quad \qquad \quad \bigg)$

\vspace*{-12ex}
\hspace*{4.5ex}
\begin{tikzpicture}
\draw [draw=red] (1.9,0) .. controls (1,0.25) and (.5,1).. (1.5,1) node at (1.5,0.5) {\tiny $p^1_1$};
\draw [draw=red] (2.1,0) .. controls (3,0.25) and (3.5,1).. (2.5,1) node at (.8,0.5) {\tiny $p^1_2$};
\draw [draw=blue] (2.5,1) .. controls (2.5,1.5) and (1.5,1.5).. (1.5,1) node at (1.9,1) {\tiny $q_1$};
\draw [draw=green] (2.5,1) .. controls (2.5,0.5) and (1.5,0.5).. (1.5,1);
\filldraw (2.5,1) circle (0.1cm);
\filldraw (1.5,1) circle (0.1cm);
\draw (2,0) circle (0.1cm);
\end{tikzpicture}
\vspace*{-9ex}
\begin{align*}
	\qquad\qquad\qquad\quad\quad\quad\quad\,\,=&\frac{\lambda^2}{(1+p^1_1+p^1_2)^2}\int_{0}^{\infty} dq_1\frac{1}{(1+q_1+p^1_1)(1+q_1+p^1_2)}\\
	=&\lambda^2\frac{\log(1+p^1_1)-\log(1+p^1_2)}{(1+p^1_1+p^1_2)^2(p^1_1-p^1_2)}
\end{align*}
\vspace*{4ex}
$\tilde{G}^{a_1^1a_2^1a_3^1}_{p_1^1p_2^1p_3^1}\,\,\Bigg(\qquad  \qquad \quad \Bigg)$

\vspace*{-12ex}
\hspace*{8ex}
\begin{tikzpicture}
\draw [draw=red] (3,0.1) .. controls (3,0.5) and (3,0.1).. (3,0.6) node at (2.3,0.2) {\tiny $p^1_1$};
\draw [draw=green] (2.9,0) .. controls (2.7,0.2) and (2.3,0.7).. (2.3,0.9) node at (2.8,0.4) {\tiny $p^1_2$};
\draw [draw=blue] (3.1,0) .. controls (3.3,0.2) and (3.7,0.7).. (3.7,0.9) node at (3.3,0.6) {\tiny $p^1_3$};
\draw [draw=red] (2.3,1.1) .. controls (2.5,1.5) and (3.5,1.5).. (3.7,1.1) node at (3,1.1) {\tiny $q_1$};
\draw [draw=blue] (2.4,1) .. controls (2.5,0.9) and (2.8,0.7).. (2.9,0.7);
\draw [draw=green] (3.6,1) .. controls (3.5,0.9) and (3.2,0.7).. (3.1,0.7);
\filldraw (3,0.7) circle (0.1cm);
\filldraw (2.3,1) circle (0.1cm);
\filldraw (3.7,1) circle (0.1cm);
\draw (3,0) circle (0.1cm);
\end{tikzpicture}
\vspace*{-10ex}
\begin{align*}
	\qquad\qquad\qquad\qquad\qquad\quad=&\frac{\lambda^3}{(1+p^1_1+p^1_2)(1+p^1_2+p^1_3)(1+p^1_1+p^1_3)}\\
	&\times \int_{0}^{\infty} dq_1\frac{1}{(1+q_1+p^1_1)(1+q_1+p^1_2)(1+q_1+p^1_3)}\\
	=&\lambda^3\frac{\frac{\log(1+p^1_1)}{(p^1_1-p^1_2)(p^1_3-p^1_1)}+\frac{\log(1+p^1_2)}{(p^1_2-p^1_1)(p^1_3-p^1_2)}+
	\frac{\log(1+p^1_3)}{(p^1_3-p^1_1)(p^1_2-p^1_3)}}{(1+p^1_1+p^1_2)(1+p^1_2+p^1_3)(1+p^1_1+p^1_3)}
\end{align*}
\vspace*{1em}

The equivalence class corresponding to the second example contains two
elements, so we obtain for $a=a^1_1=a^1_2$ 
\begin{align}\label{Loop2}
 G^{\,\,\,\,\,\,\,aa}_{2,\, p_1p_2}=\frac{2\log(\frac{1+p_1}{1+p_2})}{(1+p_1+p_2)^2(p_1-p_2)}.
\end{align}

\section{Partition function and correlation function}
In the following we demonstrate the techniques to determine
correlation functions from the partition function.  The partition
function $\mathcal{Z}[J]$ of the noncommutative 3-colour model with
external Hermitian matrices $\left(J^a_{nm}\right)$ and
$a\in\{1,2,3\}$ is formally defined by
\begin{align}\label{Zustandssumme}
	\mathcal{Z}[J]:=&\int \left(\prod_{a=1}^{3}\mathcal{D}\phi^a\right)
	\exp\left(-S[\phi]+V\sum_{a=1}^{3}\sum_{n,m=0}^\mathcal{N}J^a_{nm}\phi^a_{mn}\right)\\
	=&K\,\exp\left(-\frac{\lambda'}{3V^2}\sum_{a,b,c=1}^3\sum_{n,m,l=0}^\mathcal{N}\sigma_{abc}\frac{\partial^3}{\partial J^a_{nm}\partial J^b_{ml}\partial J^c_{ln}}
	\right)\mathcal{Z}_{free}[J],\nonumber\\
	 \mathcal{Z}_{free}[J]:=&\exp\left(\sum_{a=1}^3\sum_{n,m=0}^\mathcal{N}\frac{V}{2H_{nm}}J^a_{nm}J^a_{mn}\right),\\
	K:=&\int \left(\prod_{a=1}^{3} \mathcal{D}\phi^a\right)\exp\left(-\sum_{a=1}^3\sum_{n,m=0}^\mathcal{N}\frac{VH_{nm}}{2}\phi^a_{nm}\phi^a_{mn}\right).\nonumber
\end{align}
The logarithm of $\mathcal{Z}[J]$ will be expanded into a series of
moments with different number $B$ of boundary components.  These
moments are called correlation functions, which \textit{not
necessarily} correspond to planar graphs.  The sources are cyclic
within every boundary $\beta\in \{1,...,B\}$. For simplification we use
the notation
$\mathbb{J}_{p^\beta_1...p^\beta_{N_\beta}}^{a^\beta_1...a^\beta_{N_\beta}}:=\prod_{i=1}^{N_\beta}
J_{p_i^\beta p_{i+1}^\beta}^{a_i^\beta}$ with $N_\beta+1\equiv 1$. The
correlation functions are then defined by
\begin{align}\label{Entwicklungskoeffizienten}
	\log\frac{\mathcal{Z}[J]}{\mathcal{Z}[0]}=:\sum_{B=1}^\infty \sum_{1\leq N_1\leq...\leq N_B}^\infty \sum_{p_1^1,...,p_{N_B}^B=0}^\mathcal{N}
	\sum_{a_1^1,...,a_{N_B}^B=1}^{3}\!\!\!\!\!\!V^{2-B}\frac{G^{a_1^1...a^1_{N_1}|...|a_1^B...a^B_{N_B}}_{|p_1^1...p^1_{N_1}|...|p_1^B...p^B_{N_B}|}}{S_{(N_1,...,N_B)}}
	\prod_{\beta=1}^B\frac{\mathbb{J}_{p^\beta_1...p^\beta_{N_\beta}}^{a^\beta_1...a^\beta_{N_\beta}}}{N_\beta}.
\end{align}
If we regroup identical numbers $N_\beta$ by $(N_1,...,N_B)=(\underbrace{N_1',...,N_1'}_{\nu_1},...,\underbrace{N_s',...,N_s'}_{\nu_s})$, 
the symmetry factor is then defined by $S_{(N_1,...,N_B)}:=\prod_{i=1}^{s}\nu_i!$, due to the symmetry of boundaries with the same valence. The 
expansion coefficients $G^{a_1^1...a^1_{N_1}|...|a_1^B...a^B_{N_B}}_{|p_1^1...p^1_{N_1}|...|p_1^B...p^B_{N_B}|}$ are called 
$(N_1+...+N_B)$-point functions.
For the time being, the factor $V^{2-B}$ in 
(\ref{Entwicklungskoeffizienten}) is a matter of convention; it could
be absorbed in $G$. However, as known from \cite{Grosse2017} precisely this
convention leads to equations for the $(N_1+...+N_B)$-point functions 
which all have a well-defined large $(\mathcal{N},V)$-limit.

Due to the vanishing 1-point function for the 3-colour model, the partition function can be expanded 
with (\ref{Entwicklungskoeffizienten}) to 
\begin{align}\label{partitionexpansion}
	 &\frac{\mathcal{Z}[J]}{\mathcal{Z}[0]}=1+
	\sum_{a,b=1}^3\sum_{n,m=0}^\mathcal{N}\left(\frac{V}{2}G^{ab}_{|nm|}\mathbb{J}^{ab}_{nm}+
	\frac{1}{2}G^{a|b}_{|n|m|}\mathbb{J}^a_{n}\mathbb{J}^b_{m}\right)\\
	 &+\sum_{a,b,c=1}^3\sum_{n,m,l=0}^\mathcal{N}\Bigg(\frac{V}{3}G^{abc}_{|nml|}\mathbb{J}^{abc}_{nml}
	 +\frac{1}{2}G^{a|bc}_{|n|ml|}\mathbb{J}^a_{n}\mathbb{J}^{bc}_{ml}+\frac{1}{6V}G^{a|b|c}_{|n|m|l|}\mathbb{J}^a_{n}
	 \mathbb{J}^b_{m}\mathbb{J}^c_{l}\Bigg)\nonumber\\
	 &+\sum_{a,b,c,d=1}^3\sum_{n,m,l,p=0}^\mathcal{N}\Bigg(\frac{V}{4}G^{abcd}_{|nmlp|}\mathbb{J}^{abcd}_{nmlp}
	 +\frac{1}{3}G^{a|bcd}_{|n|mlp|}\mathbb{J}^a_{n}\mathbb{J}^{bcd}_{mlp}\nonumber\\
	 &+\left(\frac{1}{8}G^{ab|cd}_{|nm|lp|}+\frac{V^2}{8}G^{ab}_{|nm|}G^{cd}_{|lp|}\right)\mathbb{J}^{ab}_{nm}\mathbb{J}^{cd}_{lp}
	 +\left(\frac{1}{4V}G^{a|b|cd}_{|n|m|lp|}+\frac{V}{4}G^{a|b}_{|n|m|}G^{cd}_{|lp|}\right)\mathbb{J}^a_{n}
	 \mathbb{J}^b_{m}\mathbb{J}^{cd}_{lp}\nonumber\\
	 &+\left(\frac{1}{24V^2}G^{a|b|c|d}_{|n|m|l|p|}+\frac{1}{8}G^{a|b}_{|n|m|}G^{c|d}_{|l|p|}\right)\mathbb{J}^a_{n}
	 \mathbb{J}^b_{m}\mathbb{J}^c_{l}\mathbb{J}^d_{p}\Bigg)+\dots\nonumber\quad .
\end{align}

The calculation rule for later purpose is
\begin{align*}
	\frac{\partial}{\partial J^a_{p_1p_2}}J^b_{p_3p_4}=\delta_{ab}\delta_{p_1p_3}\delta_{p_2p_4}+J^b_{p_3p_4}\frac{\partial}{\partial J^a_{p_1p_2}}.
\end{align*}

\newpage
\section{Ward-Takahashi identity}
The Ward-Takahashi identity is obtained by the requirement of invariance of $\mathcal{Z}[J]$ under inner automorphisms.
For a colour model we choose a transformation as follows:
$\phi^a \mapsto (\phi^a)'=U^\dagger \phi^a U$ for $U\in \mathrm{U}(\mathcal{N})$ for \textit{one} colour $a\in \{1,2,3\}$.
The Ward-Takahashi identity following from this transformation \cite{Grossea,Disertori2007} for $p_1\neq p_2$ is given by
\begin{align}\label{Ward1}
	\sum_{m=0}^\mathcal{N}&\frac{\partial^2}{\partial J^a_{p_1m}\partial J^a_{mp_2}}\mathcal{Z}[J]+\frac{V}{E_{p_1}-E_{p_2}}
	\sum_{m=0}^\mathcal{N}\left(J^a_{p_2m}\frac{\partial}{\partial J^a_{p_1m}}-J^a_{mp_1}\frac{\partial}{\partial J^a_{mp_2}}\right)\mathcal{Z}[J]\\
	=&\frac{\lambda'}{V(E_{p_1}-E_{p_2})}\sum_{m,n=0}^\mathcal{N}\sum_{b,c=1}^3\sigma_{abc}\left(\frac{\partial^3}{\partial J^a_{p_1m}\partial J^b_{mn}\partial J^c_{np_2}}
	-\frac{\partial^3}{\partial J^b_{p_1m}\partial J^c_{mn}\partial J^a_{np_2}}
	\right)\mathcal{Z}[J].\nonumber
\end{align}
The interaction terms are certainly not invariant under the transformation of only one colour. However, the sum over all colours in (\ref{Ward1}) gives
\begin{align}\label{Ward2}
	\sum_{a=1}^3\sum_{m=0}^\mathcal{N}\frac{\partial^2}{\partial J^a_{p_1m}\partial J^a_{mp_2}}\mathcal{Z}[J]=
	\frac{V}{(E_{p_1}-E_{p_2})}\sum_{a=1}^3\sum_{m=0}^\mathcal{N}\left(J^a_{mp_1}\frac{\partial}{\partial J^a_{mp_2}}-J^a_{p_2m}\frac{\partial}{\partial J^a_{p_1m}}\right)\mathcal{Z}[J],
\end{align}
which has the usual form of a Ward-Takahashi identity. Equation (\ref{Ward2}) shows that the interaction term is invariant under the simultaneous transformation of all three colours. 

A crucial r$\hat{\text{o}}$le plays a more general identity:

\begin{proposition}\label{WardPorp2}
	Let $p_1\neq p_2$. The generalised Ward-Takahashi identity for the 3-colour matrix model with an external field
  $E$ is  
	\begin{align*}
		\sum_{m=0}^\mathcal{N}&\frac{\partial^2}{\partial J^a_{p_1m}\partial J^b_{mp_2}}\mathcal{Z}[J]+\frac{V}{E_{p_1}-E_{p_2}}
		\sum_{m=0}^\mathcal{N}\left(J^b_{p_2m}\frac{\partial}{\partial J^a_{p_1m}}-J^a_{mp_1}\frac{\partial}{\partial J^b_{mp_2}}\right)\mathcal{Z}[J]\\
		=&\frac{\lambda'}{V(E_{p_1}-E_{p_2})}\sum_{m,n=0}^\mathcal{N}\sum_{c,d=1}^3\left(\sigma_{bcd}\frac{\partial^3}{\partial J^a_{p_1m}\partial J^c_{mn}\partial J^d_{np_2}}
		-\sigma_{acd}\frac{\partial^3}{\partial J^c_{p_1m}\partial J^d_{mn}\partial J^b_{np_2}}
		\right)\mathcal{Z}[J].
	\end{align*}
	\begin{proof}
		Let $S_{int}[\phi]= V\frac{\lambda'}{3}
		\sum_{a,b,c=1}^3\sum_{n,m,l=0}^\mathcal{N}\sigma_{abc}\phi^a_{nm}
		\phi^b_{ml}\phi^c_{ln}$ be the interaction term of the action.
		Direct computation gives then
		\begin{align*}
			&\frac{E_{p_1}-E_{p_2}}{V}\sum_{m=0}^\mathcal{N}\frac{\partial^2}{\partial J^a_{p_1m}\partial J^b_{mp_2}}\mathcal{Z}[J]\\
			=&\frac{1}{V}\sum_{m=0}^\mathcal{N}\frac{\partial^2}{\partial J^a_{p_1m}\partial J^b_{mp_2}}\left((E_{p_1}+E_m)-(E_m+E_{p_2})\right)\mathcal{Z}[J]\\
			=&K\sum_{m=0}^\mathcal{N}\Bigg\{\frac{\partial}{\partial J^b_{mp_2}}\exp\left(-S_{int}\left[\frac{1}{V}\frac{\partial}{\partial J}\right]\right)J^a_{mp_1}\\
			&\qquad \qquad-
			\frac{\partial}{\partial J^a_{p_1m}}\exp\left(-S_{int}\left[\frac{1}{V}\frac{\partial}{\partial J}\right]\right)J^b_{p_2m}\Bigg\}\mathcal{Z}_{free}[J]\\
			=&\sum_{m=0}^\mathcal{N}\left(J^a_{mp_1}\frac{\partial}{\partial J^b_{mp_2}}-J^b_{p_2m}\frac{\partial}{\partial J^a_{p_1m}}\right)\mathcal{Z}[J]\\
			&-\frac{\lambda'}{V^2}\sum_{m,n=0}^\mathcal{N}\sum_{c,d=1}^3\left(\sigma_{acd}\frac{\partial^3}{\partial J^c_{p_1n}\partial J^d_{nm}\partial J^b_{mp_2}}-
			\sigma_{bcd}\frac{\partial^3}{\partial J^a_{p_1m}\partial J^c_{mn}\partial J^d_{np_2}}\right)\mathcal{Z}[J].
\end{align*}
We have used the second form of $\mathcal{Z}[J]$ in
(\ref{Zustandssumme}) and the Leibniz rule in the last
step. Technically one expands the exponential function and resums after
using the Leibniz rule. Since $E_{p_1}\neq E_{p_2}$ the proof is
finished.
	\end{proof}
\end{proposition}

Equation (\ref{Ward1}) is a special case of Proposition
\ref{WardPorp2} by setting $b=a$.  The derivation of both identities
is completely different. Proposition \ref{WardPorp2} can not be obtained 
by a symmetry transformation of only one colour
due to the discrete mixing of the colours if $a\neq b$.
Applying the procedure of the proof of Proposition \ref{WardPorp2}, it
is also possible to derive the usual Ward-Takahashi identity even in
other models.

For later purpose we combine two identities to get a more useful expression:
\begin{lemma}\label{WarsLemma}
	Let $a$ be fixed and $p_1\neq p_2$, then it follows
	\begin{align*}
		&\sum_{b,c=1}^{3}\sum_{m=0}^\mathcal{N}\sigma_{abc}\frac{\partial^2}{\partial J^b_{p_1m}\partial J^c_{mp_2}}\mathcal{Z}[J]
		\\
		=&\frac{V}{E_{p_1}-E_{p_2}}\Bigg[\sum_{b,c=1}^{3}\sigma_{abc}
		\sum_{m=0}^\mathcal{N}\left(J^b_{mp_1}\frac{\partial}{\partial J^c_{mp_2}}-J^c_{p_2m}\frac{\partial}{\partial J^b_{p_1m}}\right)\\
		&\qquad\qquad+\frac{\lambda'}{V^2}\sum_{b=1}^3\Bigg\{\sum_{m=0}^\mathcal{N}\left(\frac{\partial^3}{\partial J^b_{p_1m}\partial J^b_{mp_1}\partial J^a_{p_1p_2}}-\frac{\partial^3}{\partial J^a_{p_1p_2}\partial J^b_{p_2m}\partial J^b_{mp_2}}\right)\\
		&\qquad\qquad+\sum_{\substack{m,n=0 \\ n\neq p_1}}^{\mathcal{N}}\frac{V}{E_{p_1}-E_n}\frac{\partial}{\partial J^a_{np_2}}\left(J^b_{mp_1}\frac{\partial}{\partial J^b_{mn}}-J^b_{nm}\frac{\partial}{\partial J^b_{p_1m}}\right)\\
		&\qquad\qquad-\sum_{\substack{m,n=0 \\ n\neq p_2}}^{\mathcal{N}}\frac{V}{E_{p_2}-E_{n}}\frac{\partial}{\partial J^a_{p_1n}}\left(J^b_{p_2m}\frac{\partial}{\partial J^b_{nm}}-J^b_{mn}\frac{\partial}{\partial J^b_{mp_2}}\right)\Bigg\}\Bigg]\mathcal{Z}[J].
	\end{align*}
	\begin{proof}
Inserting Proposition \ref{WardPorp2} for the LHS yields 
		\begin{align}\label{GLLemmaWard}
			&\sum_{b,c=1}^{3}\sum_{m=0}^\mathcal{N}\sigma_{abc}\frac{\partial^2}{\partial J^b_{p_1m}\partial J^c_{mp_2}}\mathcal{Z}[J]\nonumber\\
			=&\frac{V}{E_{p_1}-E_{p_2}}\sum_{b,c=1}^{3}\sigma_{abc}
			\sum_{m=0}^\mathcal{N}\left(J^b_{mp_1}\frac{\partial}{\partial J^c_{mp_2}}-J^c_{p_2m}\frac{\partial}{\partial J^b_{p_1m}}\right)\mathcal{Z}[J]\\
			+&\frac{\lambda'}{V(E_{p_1}-E_{p_2})}\!\!\sum_{m,n=0}^\mathcal{N}\sum_{b,c,d,e=1}^3\!\!\!\!\!\sigma_{abc}
			\!\left(\! \sigma_{cde}\frac{\partial^3}{\partial J^b_{p_1m}\partial J^d_{mn}\partial J^e_{np_2}}-\sigma_{bde}\frac{\partial^3}{\partial J^d_{p_1m}\partial J^e_{mn}\partial J^c_{np_2}}\nonumber
			\right)\!\mathcal{Z}[J].
\end{align} 
By the sum over the colours $b,c,d,e$, we obtain for the
multiplication of two $\sigma$'s with one common index
\begin{align*}
			\sigma_{abc}\sigma_{cde}=&\sigma_{abc}(\delta_{ad}\delta_{be}+\delta_{ae}\delta_{bd})\\
			\sigma_{abc}\sigma_{bde}=&\sigma_{abc}(\delta_{ad}\delta_{ce}+\delta_{ae}\delta_{cd}).
		\end{align*}
Therefore, the last line in (\ref{GLLemmaWard}) gives 
\begin{align}\label{proofGL1}
			&\frac{\lambda'}{V(E_{p_1}-E_{p_2})}\sum_{m,n=0}^\mathcal{N}\sum_{b,c=1}^3\sigma_{abc}\Bigg(\frac{\partial^3}{\partial J^b_{p_1m}\partial J^a_{mn}\partial J^b_{np_2}}+\frac{\partial^3}{\partial J^b_{p_1m}\partial J^b_{mn}\partial J^a_{np_2}}\\
			&\qquad\qquad\qquad\qquad\qquad\qquad\qquad\quad-\frac{\partial^3}{\partial J^a_{p_1m}\partial J^c_{mn}\partial J^c_{np_2}}-\frac{\partial^3}{\partial J^c_{p_1m}\partial J^a_{mn}\partial J^c_{np_2}}\Bigg)\mathcal{Z}[J].\nonumber
\end{align}
The first and the last term in parentheses vanish because of the
total symmetry of $\sigma_{abc}$. Adding
$0=\left(\frac{\partial^3}{\partial J^a_{p_1m}\partial
    J^a_{mn}\partial J^a_{np_2}}-\frac{\partial^3}{\partial
    J^a_{p_1m}\partial J^a_{mn}\partial
    J^a_{np_2}}\right)\mathcal{Z}[J]$ and renaming the indices,
(\ref{proofGL1}) can be rewritten to
\begin{align*}
			\frac{\lambda'}{V(E_{p_1}-E_{p_2})}\sum_{m,n=0}^\mathcal{N}\sum_{b=1}^3\left(\frac{\partial^3}{\partial J^b_{p_1m}\partial J^b_{mn}\partial J^a_{np_2}}-\frac{\partial^3}{\partial J^a_{p_1m}\partial J^b_{mn}\partial J^b_{np_2}}\right)\mathcal{Z}[J].
		\end{align*}
		Inserting (\ref{Ward2}) for $n\neq p_1$ in the first and $m\neq p_2$ in the second term gives after renaming indices finally
		\begin{align}\label{GLLemmaWard2}
			\frac{\lambda'}{V(E_{p_1}-E_{p_2})}\sum_{b=1}^3&\Bigg\{\sum_{m=0}^\mathcal{N}\left(\frac{\partial^3}{\partial J^b_{p_1m}\partial J^b_{mp_1}\partial J^a_{p_1p_2}}-\frac{\partial^3}{\partial J^a_{p_1p_2}\partial J^b_{p_2m}\partial J^b_{mp_2}}\right)\nonumber\\*
			&+\sum_{\substack{m,n=0 \\ n\neq p_1}}^{\mathcal{N}}\frac{V}{E_{p_1}-E_n}\frac{\partial}{\partial J^a_{np_2}}\left(J^b_{mp_1}\frac{\partial}{\partial J^b_{mn}}-J^b_{nm}\frac{\partial}{\partial J^b_{p_1m}}\right)\\*
			&-\sum_{\substack{m,n=0 \\ n\neq p_2}}^{\mathcal{N}}\frac{V}{E_{p_2}-E_{n}}\frac{\partial}{\partial J^a_{p_1n}}\left(J^b_{p_2m}\frac{\partial}{\partial J^b_{nm}}-J^b_{mn}\frac{\partial}{\partial J^b_{mp_2}}\right)\Bigg\}\mathcal{Z}[J]\nonumber.
		\end{align}
		The identity follows by combining (\ref{GLLemmaWard}) and (\ref{GLLemmaWard2}).
	\end{proof}
\end{lemma}

\section{Schwinger-Dyson equations for $B=1$}
\subsection{For matrix basis}
In this section we derive the Schwinger-Dyson equations with the help
of Ward-Takahashi identity.
\begin{proposition}\label{Prop2Punkt}
 The Schwinger-Dyson equation for the 2-point function in the 3-colour matrix model with an external field
 $E$ is for $p_1\neq p_2$ given by
 \begin{align*}
  &G^{aa}_{|p_1p_2|}=\frac{1}{H_{p_1p_2}}+\frac{\lambda'^2}{(E^2_{p_1}-E^2_{p_2})V}\\
  &\times\Bigg[\sum_{m=0}^\mathcal{N}\sum_{b=1}^3\left(G^{aa}_{|p_1p_2|}\left(
 G^{bb}_{|p_2m|}
-G^{bb}_{|p_1m|}\right)+\frac{1}{V}\left(G^{aabb}_{|p_2p_1p_2m|}-G^{aabb}_{|p_1p_2p_1m|}\right)\right)\\
&+\sum_{b=1}^3\frac{1}{V^2}\bigg(\sum_{m=0}^\mathcal{N}\left(G^{aa|bb}_{|p_2p_1|p_2m|}-G^{aa|bb}_{|p_1p_2|p_1m|}\right)
+\left(G^{b|baa}_{|p_2|p_2p_2p_1|}-G^{b|baa}_{|p_1|p_1p_1p_2|}\right)\bigg)\\
&+\sum_{b=1}^3\left(\frac{1}{V^3}\left(G^{b|b|aa}_{|p_2|p_2|p_2p_1|}-G^{b|b|aa}_{|p_1|p_1|p_1p_2|}\right)
+\frac{1}{V}G^{aa}_{|p_1p_2|}\left(G^{b|b}_{|p_2|p_2|}-G^{b|b}_{|p_1|p_1|}\right)\right)\\
&+\sum_{\substack{m=0\\ m\neq p_2}}^\mathcal{N}\frac{G^{aa}_{|p_1m|}-G^{aa}_{|p_1p_2|}}{E_{p_2}-E_m}-
\sum_{\substack{m=0\\ m\neq p_1}}^\mathcal{N}
\frac{G^{aa}_{|p_1p_2|}-G^{aa}_{|p_2m|}}{E_{m}-E_{p_1}}+\frac{1}{V}\frac{G^{a|a}_{|p_1|p_1|}-2G^{a|a}_{|p_1|p_2|}+G^{a|a}_{|p_2|p_2|}}{E_{p_2}-E_{p_1}}\Bigg].
 \end{align*}
\begin{proof}
 Assuming $p_1\neq p_2$ the 2-point function is given via 
definition (\ref{Entwicklungskoeffizienten}) and expansion
 (\ref{partitionexpansion}). Using (\ref{Zustandssumme}) leads to
 \begin{align*}
  G^{aa}_{|p_1p_2|}
  =&\frac{1}{V}\frac{\partial^2}{\partial J^a_{p_1p_2}\partial J^a_{p_2p_1}}\mathrm{log}\mathcal{Z}[J]\Big\vert_{J=0}
  =\frac{1}{V\mathcal{Z}[0]}\frac{\partial^2}{\partial J^{a}_{p_1p_2}\partial J^{a}_{p_2p_1}}\mathcal{Z}[J]\Big|_{J=0}\\
  =&\frac{K}{H_{p_1p_2}\mathcal{Z}[0]}\frac{\partial}{\partial J^{a}_{p_2p_1}}\exp\left(-S_{int}\left[\frac{1}{V}\frac{\partial}{\partial J}\right]\right)
  J^{a}_{p_2p_1}\mathcal{Z}_{free}[J]\Big|_{J=0}\\
  =&\frac{1}{H_{p_1p_2}}-\frac{\lambda' }{H_{p_1p_2}\mathcal{Z}[0]V^2}\frac{\partial}{\partial J^{a}_{p_2p_1}}
 \sum_{b,c=1}^3\sum_{m=0}^\mathcal{N}\sigma_{abc}\frac{\partial^2}{\partial J^b_{p_1m}\partial J^c_{mp_2}}\mathcal{Z}[J]\Big|_{J=0}.
 \end{align*}
 Inserting the expansion of (\ref{partitionexpansion}) would give the
 Schwinger-Dyson equation between the 2-point and 3-point function.
 At first sight the application of Lemma
 \ref{WarsLemma} seems to make the equation more complicated.  However,
 it yields a better behaviour in the large ($\mathcal{N},V$)-limit. The
 first term on the RHS of the equation of Lemma \ref{WarsLemma}
 vanishes by setting $J$ to zero.  Therefore, we obtain
 \begin{align*}
  =&\frac{1}{H_{p_1p_2}}-\frac{\lambda'^2 }{(E^2_{p_1}-E^2_{p_2})\mathcal{Z}[0]V^3}\\
  &\times\Bigg\{
  \sum_{b=1}^3\sum_{m=0}^\mathcal{N}\left(\frac{\partial^4}{\partial J^b_{p_1m}\partial J^b_{mp_1}\partial J^a_{p_1p_2}\partial J^a_{p_2p_1}}-
  \frac{\partial^4}{\partial J^a_{p_2p_1}\partial J^a_{p_1p_2}\partial J^b_{p_2m}\partial J^b_{mp_2}}\right)\mathcal{Z}[J]\Big|_{J=0}\\
  &\qquad+\sum_{\substack{m,n=0 \\ n\neq p_1}}^{\mathcal{N}}\frac{V}{E_{p_1}-E_n}\frac{\partial^2}{\partial J^a_{p_2p_1}\partial J^a_{np_2}}\left(J^a_{mp_1}\frac{\partial}{\partial J^a_{mn}}-
  J^a_{nm}\frac{\partial}{\partial J^a_{p_1m}}\right)\mathcal{Z}[J]\Big|_{J=0}\\
  &\qquad-\sum_{\substack{m,n=0 \\ n\neq p_2}}^{\mathcal{N}}\frac{V}{E_{p_2}-E_{n}}\frac{\partial^2}{\partial J^a_{p_2p_1}\partial J^a_{p_1n}}\left(J^a_{p_2m}\frac{\partial}{\partial J^a_{nm}}
  -J^a_{mn}\frac{\partial}{\partial J^a_{mp_2}}\right)
  \mathcal{Z}[J]\Big|_{J=0}\Bigg\},
 \end{align*}
  where $H_{p_1p_2}(E_{p_1}-E_{p_2})=(E^2_{p_1}-E^2_{p_2})$ has been used and the fact that in the last two lines only colour $a$ survives. 
  By taking $p_1\neq p_2$ into account and $J=0$ gives with the Leibniz rule
  \begin{align*}
    =&\frac{1}{H_{p_1p_2}}-\frac{\lambda'^2 }{(E^2_{p_1}-E^2_{p_2})\mathcal{Z}[0]V^3}\\
  &\times\Bigg\{
  \sum_{b=1}^3\sum_{m=0}^\mathcal{N}\left(\frac{\partial^4}{\partial J^b_{p_1m}\partial J^b_{mp_1}\partial J^a_{p_1p_2}\partial J^a_{p_2p_1}}-
  \frac{\partial^4}{\partial J^a_{p_2p_1}\partial J^a_{p_1p_2}\partial J^b_{p_2m}\partial J^b_{mp_2}}\right)\mathcal{Z}[J]\Big|_{J=0}\\
  &\qquad+\sum_{\substack{m=0 \\ m\neq p_1}}^{\mathcal{N}}\frac{V}{E_{p_1}-E_m}
  \left(\frac{\partial^2}{\partial J^a_{mp_2}\partial J^a_{p_2m}}-
  \frac{\partial^2}{\partial J^a_{p_1p_2}\partial J^a_{p_2p_1}}\right)\mathcal{Z}[J]\Big|_{J=0}\\
  &\qquad-\sum_{\substack{m=0 \\ m\neq p_2}}^{\mathcal{N}}\frac{V}{E_{p_2}-E_{m}}
  \left(\frac{\partial^2}{\partial J^a_{mp_1}\partial J^a_{p_1m}}-
  \frac{\partial^2}{\partial J^a_{p_1p_2}\partial J^a_{p_2p_1}}\right)\mathcal{Z}[J]\Big|_{J=0}\\
  &\qquad+ \frac{V}{E_{p_2}-E_{p_1}}\left( \frac{\partial^2}{\partial J^a_{p_2p_2}\partial J^a_{p_1p_1}}-
  \frac{\partial^2}{\partial J^a_{p_2p_2}\partial J^a_{p_1p_1}}\right)\mathcal{Z}[J]\Big|_{J=0}\Bigg\}.
  \end{align*}
  The first line generates for $m\neq p_1$ and $m\neq p_2$ either a 4-point function with one boundary or two 2-point functions with one boundary, respectively.
  Functions with higher boundaries $B\geq 2$ appear in case of $m=p_1$ or $m=p_2$ . All terms are found by comparing with the expansion
  (\ref{partitionexpansion}).
\end{proof}
\end{proposition}

We remind that in Proposition \ref{Prop2Punkt} correlation functions
of genus $g\geq1$ are also included. To see this 
one has to expand the correlation functions in a genus expansion. More information can be found in \cite{Grossea}. 
The Schwinger-Dyson equation of the 2-point function depends on $\lambda'^2$, since graphs exist only with an even number of vertices.

\begin{proposition}\label{PropNPunkt}
 Let $N\geq 3$. The Schwinger-Dyson equation for the $N$-point function in the 3-colour matrix model with an external field
 $E$ is for pairwise different $p_i, p_j$ given by 
 \begin{align*}
  &G^{a_1...a_N}_{|p_1...p_N|}
  =-\frac{\lambda' }{(E^2_{p_1}-E^2_{p_2})}\sum_{b=1}^3\left(\sigma_{a_1a_Nb}G^{a_2...a_{N-1}b}_{|p_2...p_{N-1}p_N|}
 -\sigma_{a_1a_2b}G^{ba_3a_4...a_{N}}_{|p_1p_3p_4...p_N|}\right)\nonumber\\
 &-\frac{\lambda'^2 }{V^2(E^2_{p_1}-E^2_{p_2})}\\
 &\times\Bigg\{V\Bigg(\sum_{\substack{m=0\\m\neq p_1}}^\mathcal{N}\frac{G^{a_1a_2...a_N}_{|mp_2...p_N|}-G^{a_1a_2...a_N}_{|p_1p_2...p_N|}}{E_{p_1}-E_m}
 -\sum_{\substack{m=0\\m\neq p_2}}^\mathcal{N}\frac{G^{a_1a_2a_3...a_N}_{|p_1mp_3...p_N|}-G^{a_1a_2...a_N}_{|p_1p_2...p_N|}}{E_{p_2}-E_m}\Bigg)\nonumber\\
&+\sum_{k=2}^{N}\Bigg(\frac{G^{a_1a_2...a_{k-1}|a_ka_{k+1}...a_N}_{|p_kp_2...p_{k-1}|p_kp_{k+1}...p_N|}-G^{a_1a_2...a_{k-1}|a_ka_{k+1}...a_N}_{|p_kp_2...p_{k-1}|p_1p_{k+1}...p_N|}}{E_{p_1}-E_{p_k}}\\
&\qquad\qquad\qquad-\frac{G^{a_2a_3...a_{k}|a_1a_{k+1}...a_N}_{|p_{k+1}p_3...p_{k}|p_1p_{k+1}...p_N|}-
G^{a_2...a_k|a_1a_{k+1}...a_N}_{|p_2...p_k|p_1p_{k+1}...p_N|}}{E_{p_2}-E_{p_{k+1}}}\Bigg)\nonumber\\
&+\sum_{k=3}^{N-1}V^2\Bigg(G^{a_1a_2...a_{k-1}}_{|p_kp_2...p_{k-1}|}\frac{G^{a_k...a_N}_{|p_k...p_N|}
-G^{a_ka_{k+1}...a_N}_{|p_1p_{k+1}...p_N|}}{E_{p_1}-E_{p_k}}\\
&\qquad\qquad\qquad-G^{a_1a_{k+1}...a_N}_{|p_1p_{k+1}...p_N|}\frac{G^{a_2a_3...a_{k}}_{|p_{k+1}p_3...p_{k}|}
-G^{a_2...a_k}_{|p_2...p_k|}}{E_{p_2}-E_{p_{k+1}}}\Bigg)\\
&+\sum_{b=1}^3\sum_{m=0}^\mathcal{N}\bigg(G^{bba_1...a_N}_{|p_1mp_1...p_N|}-G^{a_1bba_2...a_N}_{|p_1p_2mp_2...p_N|}+\frac{1}{V}
\left(G^{bb|a_1...a_N}_{|p_1m|p_1...p_N|}-G^{bb|a_1...a_N}_{|p_2m|p_1...p_N|}\right)\nonumber\\
&\qquad\qquad\qquad+VG^{a_1...a_N}_{|p_1...p_N|}\left(G^{bb}_{|p_1m|}-G^{bb}_{|p_2m|}\right)\bigg)\nonumber\\
&+\sum_{b=1}^3\sum_{k=2}^N\bigg(\frac{1}{V}\left(G^{ba_1...a_{k-1}|ba_k...a_N}_{|p_kp_1...p_{k-1}|p_1p_k...p_N|}-G^{ba_2...a_{k}|ba_{k+1}...a_Na_1}_{|p_{k+1}p_2...p_{k}|p_2p_{k+1}...p_Np_1|}\right)\\
&\qquad\qquad\qquad+V\left(G^{ba_1...a_{k-1}}_{|p_kp_1...p_{k-1}|}G^{ba_k...a_N}_{|p_1p_k...p_N|}-G^{ba_2...a_{k}}_{|p_{k+1}p_2...p_{k}|}G^{ba_{k+1}...a_Na_1}_{|p_2p_{k+1}...p_Np_1|}\right)\bigg)\\
&+\sum_{b=1}^3\bigg(\frac{1}{V^2}\left(G^{b|b|a_1...a_N}_{|p_1|p_1|p_1...p_N|}-G^{b|b|a_1...a_N}_{|p_2|p_2|p_1...p_N|}\right)+
\frac{1}{V}\left(G^{b|ba_1...a_N}_{|p_1|p_1p_1...p_N}-G^{b|ba_2...a_Na_1}_{|p_2|p_2p_2...p_Np_1}\right)\\*
&\qquad\qquad\qquad+G^{a_1...a_N}_{|p_1...p_N|}\left(G^{b|b}_{|p_1|p_1|}-G^{b|b}_{|p_2|p_2|}\right) \bigg)  \Bigg\},
 \end{align*}
 where $p_{N+1}\equiv p_1$.
\begin{proof}
 We use the definition of the $N$-point function for pairwise different $p_i,p_j$. With the expression of the partition function
 (\ref{Zustandssumme}), we obtain
 \begin{align*}
  G^{a_1...a_N}_{|p_1...p_N|}=&\frac{1}{V}\frac{\partial^N}{\partial J^{a_1}_{p_1p_2}...J^{a_N}_{p_Np_1}}\frac{\mathcal{Z}[J]}{\mathcal{Z}[0]}\bigg|_{J=0}\nonumber\\
=&-\frac{\lambda' }{H_{p_1p_2}V^2\mathcal{Z}[0]}\frac{\partial^{N-1}}{\partial J^{a_2}_{p_2p_3}...J^{a_N}_{p_Np_1}}
\sum_{b,c=1}^3\sigma_{a_1bc}\sum_{n=0}^\mathcal{N}\frac{\partial^{2}}{\partial J^{b}_{p_1n}J^{c}_{np_2}}\mathcal{Z}[J]\bigg|_{J=0}.
 \end{align*}
Here the first derivative $\frac{\partial}{\partial J^{a_1}_{p_1p_2}}$
  applied to $\mathcal{Z}_{free}[J]$ yields $\frac{V}{H_{p_1p_2}}
J^{a_1}_{p_2p_1}$, which can only be differentiated by the interaction
in $\mathcal{Z}[Z]$ because of $p_3\neq p_1$ and $p_2\neq p_4$.
 Applying Lemma \ref{WarsLemma} yields
 \begin{subequations}
 \begin{align}
  =&-\frac{\lambda' }{(E^2_{p_1}-E^2_{p_2})V\mathcal{Z}[0]}\frac{\partial^{N-1}}{\partial J^{a_2}_{p_2p_3}...J^{a_N}_{p_Np_1}}\nonumber\\
  &\times \Bigg[\sum_{b,c=1}^{3}\sigma_{a_1bc}
		\sum_{m=0}^\mathcal{N}\left(J^b_{mp_1}\frac{\partial}{\partial J^c_{mp_2}}-J^c_{p_2m}\frac{\partial}{\partial J^b_{p_1m}}\right)\label{I1}\\
		&\qquad\qquad+\frac{\lambda'}{V^2}\sum_{b=1}^3\Bigg\{\sum_{m=0}^\mathcal{N}\left(\frac{\partial^3}{\partial J^b_{p_1m}\partial J^b_{mp_1}\partial J^{a_1}_{p_1p_2}}
		-\frac{\partial^3}{\partial J^{a_1}_{p_1p_2}\partial J^b_{p_2m}\partial J^b_{mp_2}}\right)\label{I2}\\
		&\qquad\qquad+\sum_{\substack{m,n=0 \\ n\neq p_1}}^{\mathcal{N}}\frac{V}{E_{p_1}-E_n}\frac{\partial}
		{\partial J^{a_1}_{np_2}}\left(J^b_{mp_1}\frac{\partial}{\partial J^b_{mn}}-J^b_{nm}\frac{\partial}{\partial J^b_{p_1m}}\right)\label{I3}\\
		&\qquad\qquad-\sum_{\substack{m,n=0 \\ n\neq p_2}}^{\mathcal{N}}\frac{V}{E_{p_2}-E_{n}}\frac{\partial}{\partial J^{a_1}_{p_1n}}
		\left(J^b_{p_2m}\frac{\partial}{\partial J^b_{nm}}-J^b_{mn}\frac{\partial}{\partial J^b_{mp_2}}\right)\Bigg\}\Bigg]\mathcal{Z}[J]\bigg|_{J=0}\label{I4}.
 \end{align}
 \end{subequations}
The first term of (\ref{I1}) contributes only for $b=a_N$ and $m=p_N$ and the second term only for $c=a_2$ and $m=p_3$. This generates
the term proportional to $\lambda'$. Line (\ref{I2}) produces three 
different types of terms for arbitrary $m$, the $(2+N)$-point functions with $B=1$, the multiplication of 2-point 
with $N$-point functions, and $(2+N)$-point functions with $B=2$. 
If in (\ref{I2}) $m=p_k$ for the first term with $2\leq k\leq N$ (for the second term with $3\leq k\leq N$ or $k=1$), additionally 
$(k+(N+2-k))$-point 
functions with $B=2$ and the multiplication of $k$-point with $(N+2-k)$-point functions with $B=1$ are generated.
In case of $m=p_1$ for the left term ($m=p_2$ for the right term)  (\ref{I2}) produces either $(1+1+N)$-point functions with $B=3$, $(1+(1+N))$-point functions with $B=2$
or the multiplication of $(1+1)$-point with $N$-point functions. 

Finally, we look at (\ref{I3}) and (\ref{I4}) together. The first terms
again contribute only for $b=a_N$ and $m=p_N$ in (\ref{I3}) or for $b=a_2$ and $m=p_3$ in (\ref{I4}). Since the sum over $n$ survives, 
$N$-point functions arise. If $n=p_k$ for $k\neq 1$ in (\ref{I3}) and for $k\neq 2$ in (\ref{I4}) one gets either $(k+(N-k))$-point functions or 
the multiplication of $k$-point functions with $(N-k)$-point functions with $B=1$. For 
the second term in (\ref{I3}) and (\ref{I4}) each derivative have to be taken into account.
If the derivative in front of the brackets in (\ref{I3}) and (\ref{I4}) acts on $J^b_{nm}$ or $J^b_{mn}$, the sum over $n$ survives again
and has a prefactor depending on $E_n$, but no $n$ appears in the $N$-point function. 
If any other derivative $\frac{\partial}{\partial J^{a_{k+1}}_{p_{k+1} p_{k+2}}}$, for some $k\geq 1$, acts on the second term, $n,m,b$ will be
fixed and it will produces $N$-point functions, $(k+(N-k))$-point functions with $B=2$ and the multiplication of $k$-point with
$(N-k)$-point functions. Collecting all and making use of (\ref{Entwicklungskoeffizienten}) to get the correct prefactor in
$V$, one find all the terms appearing in Proposition \ref{PropNPunkt}.
\end{proof}
\end{proposition}
The first term shows that a $(N-1)$-point function only contributes for different adjacent colours, because of $\sigma_{a_1a_Nb}$ and 
$\sigma_{a_1a_2b}$. This fact fits perfectly with a loop expansion. Furthermore, the 2-point function is assigned with a special 
r$\hat{\text{o}}$le, since the sum over $m$ only appears for the $N$-point and 2-point function even in the large $\mathcal{N},V$
limit.

It should be emphasised that not all combinations of the colours for the correlation functions are possible. The 2-point function is of the form $G^{aa}_{|p_1p_2|}$ and
the 3-point function $\sigma_{abc}G^{abc}_{|p_1p_2p_3|}$. 
There exists no 4-point function equipped with all three colours simultaneously, and so on. These properties which are first 
recognized by loop expansion are intrinsically presented in the Schwinger-Dyson equations. 

\subsection{Large ($\mathcal{N},V$)-limit}
Sending $\mathcal{N},V \to \infty$ with constant ratio $\frac{\mathcal{N}}{V}=\mu^2\Lambda^2$, the sum is turned into an integral by the 
transformation of the discrete elements to continuous variables $m \to V\mu^2 q$
\begin{align*}
 \lim_{\substack{\mathcal{N},V\to\infty\\ \frac{\mathcal{N}}{V}=\mu^2\Lambda^2}}\frac{1}{V}\sum_{m=0}^\mathcal{N}f\left(\frac{m}{V}\right)=
 \mu^2\int_0^{\Lambda^2}dq\, f(\mu^2q).
\end{align*}
The eigenvalues of the external field are in the linear case given by $E_m=\mu^2(q+\frac{1}{2})$.
To get rid of the mass squared $\mu^2$, we redefine in the following way
\begin{align*}
  G^{aa}_{p_1p_2}:=\lim_{\mathcal{N},V\to\infty}\mu^2 G^{aa}_{|p_1p_2|},\qquad\lambda:=\frac{\lambda'}{\mu^2}.
\end{align*}
An important fact is that in this limit only functions with genus $g=0$ survive \cite{Grossea}. Furthermore, we assume that all 
these functions of genus zero are $\mathcal{O}(V^0)$. The perturbation theory shows that in combination with
the definition (\ref{Entwicklungskoeffizienten}) this is the right 
assumption (see also \cite{Grosse2017,Grosse2018}).

The equation of Proposition 
\ref{Prop2Punkt} breaks down to a closed equation, since 4-point functions and functions with $B\geq 2$ vanish. 
The limit $\lim_{p_2\to p_1} \frac{G^{\,\,\,\,aa}_{n,p_1q}-G^{\,\,\,\,aa}_{n,p_2q}}{p_1-p_2}$ is in perturbation theory well-defined, 
therefore this limit should also exist in the non-perturbative case. 
Sending $\Lambda^2\to \infty$ the closed integral equation for 
the 3-colour model is obtained
\begin{align}\label{GeschlosseneGL}
 G^{aa}_{p_1p_2}=&
\frac{1}{1+p_1+p_2}+
  \frac{\lambda^2}{\left(1+p_1+p_2\right)(p_1-p_2)}
  \bigg(3G^{aa}_{p_1p_2}\int_0^{\infty}dq \,\left(G^{aa}_{qp_2}-G^{aa}_{p_1q}\right)\\
&\qquad\qquad\qquad\qquad-\int_0^{\infty}dq\,\frac{G^{aa}_{p_1q}
-G^{aa}_{p_1p_2}}{q-p_2}+\int_0^{\infty}dq\,
\frac{G^{aa}_{p_2q}-G^{aa}_{p_1p_2}}{q-p_1}\bigg).\nonumber
\end{align}
We have assumed that $G^{bb}_{p_1p_2}$ does not depend directly on the
colour $b$ so that $\sum_{b=1}^3G^{bb}_{p_1p_2}=3G^{aa}_{p_1p_2}$.

\subsection{Perturbative solution}
Using the expansion (\ref{Gloopexpansion}) the closed integral
equation (\ref{GeschlosseneGL}) provides a recursive equation for
$n\geq 1$ of the form
\begin{align}\label{rekursiveGL}
 G^{\quad\,aa}_{2n,\,p_1p_2}=&
  \frac{1}{\left(1+p_1+p_2\right)(p_1-p_2)}
  \bigg(3\sum_{i=0}^{n-1}G^{\qquad\,\,\,\quad aa}_{2(n-1-i),\,p_1p_2}\int_0^{\infty}dq \,\left(G^{\quad aa}_{2i,\,qp_2}-
  G^{\quad aa}_{2i,\,p_1q}\right)\\
&-\int_0^{\infty}dq\,\frac{G^{\qquad aa}_{2n-2,\,p_1q}
-G^{\qquad aa}_{2n-2,\,p_1p_2}}{q-p_2}+\int_0^{\infty}dq\,
\frac{G^{\qquad aa}_{2n-2,\,p_2q}-G^{\qquad aa}_{2n-2,\,p_1p_2}}{q-p_1}\bigg)\nonumber
\end{align}
and $G^{\quad aa}_{0,\,p_1p_2}=\frac{1}{1+p_1+p_2}$. Equation (\ref{rekursiveGL}) is linear which enables a very easy way to study this model 
in comparison to other noncommutative 
quantum field theory models. The convergence of loop expansion is hopeless, since the number of graphs of order $\lambda^{2n}$ is at least of
order $\mathcal{O}(n!)$, however the recursive equation gives directly the sum over all graphs of a certain order $\lambda^{2n}$.

\addsec{\large{Order $n=1$}}
It is easy to verify with $\frac{\frac{1}{1+x+y}-\frac{1}{1+x+z}}{y-z}=-\frac{1}{(1+x+y)(1+x+z)}$
the result of (\ref{Loop2}) by
\begin{align*}
 G^{\quad aa}_{2,\,p_1p_2}=&\frac{1}{\left(1+p_1+p_2\right)(p_1-p_2)}\Bigg(\frac{3}{1+p_1+p_2}\int_0^\infty 
 dq \left(\frac{1}{1+p_2+q}-\frac{1}{1+p_1+q}\right)\\*
 &\qquad\qquad\qquad\qquad-\frac{1}{1+p_1+p_2}\int_0^\infty 
 dq \left(\frac{1}{1+p_2+q}-\frac{1}{1+p_1+q}\right)\Bigg)\\*
 =&\frac{2\log(\frac{1+p_1}{1+p_2})}{(1+p_1+p_2)^2(p_1-p_2)}.
\end{align*}

\addsec{\large{Order $n=2$}}
Inserting $G^{\quad aa}_{0,\,p_1p_2}$ and $G^{\quad aa}_{2,\,p_1p_2}$ into (\ref{rekursiveGL}) to obtain
\begin{subequations}
\begin{align}
 &G^{\quad aa}_{4,\,p_1p_2}=	\frac{6\log(\frac{1+p_1}{1+p_2})}{(1+p_1+p_2)^2(p_1-p_2)^2}\int_0^\infty dq\,\frac{p_1-p_2}{(1+p_1+q)(1+p_2+q)}\label{I1'}\\
&+ \frac{6}{(1+p_1+p_2)^2(p_1-p_2)}\int_0^\infty\!\! dq\left(\frac{\log(\frac{1+p_2}{1+q})}{(1+q+p_2)^2(p_2-q)}-
 \frac{\log(\frac{1+p_1}{1+q})}{(1+q+p_1)^2(p_1-q)}\right)\label{I2'}\\
 &-\frac{2}{(1+p_1+p_2)(p_1-p_2)}\int_0^\infty dq\frac{\frac{\log(\frac{1+p_1}{1+q})}{(1+q+p_1)^2(p_1-q)}-\frac{\log(\frac{1+p_1}{1+p_2})}{(1+p_1+p_2)^2(p_1-p_2)}}{q-p_2}\label{I3'}\\
 &+\frac{2}{(1+p_1+p_2)(p_1-p_2)}\int_0^\infty dq\frac{\frac{\log(\frac{1+q}{1+p_2})}{(1+q+p_2)^2(q-p_2)}-\frac{\log(\frac{1+p_1}{1+p_2})}{(1+p_1+p_2)^2(p_1-p_2)}}{q-p_1}\label{I4'}.
\end{align}
\end{subequations}
(\ref{I1'}) is given by
\begin{align*}
 \frac{6\log(\frac{1+p_1}{1+p_2})^2}{(1+a+b)^3(a-b)^2}.
\end{align*}
Using the definition of the dilogarithm
\begin{align*}
	\mathrm{Li}_2(-x)=-\int_{0}^{x}du\frac{\log(1+u)}{u},
\end{align*} 
(\ref{I2'}) is then determined by
\begin{align*}
 \frac{6}{(1+p_1+p_2)^2(p_1-p_2)}&\Bigg(\frac{\log(1+p_1)}{p_1(1+p_1)(1+2p_1)}+\frac{\log(1+p_1)^2+2\mathrm{Li}_2\left(-p_1\right) 
 -\frac{\pi^2 }{6}}{(1+2p_1)^2}\\*
 &-\frac{\log(1+p_2)}{p_2(1+p_2)(1+2p_2)}-\frac{\log(1+p_2)^2+2\mathrm{Li}_2\left(-p_2\right) -\frac{\pi^2 }{6}}{(1+2p_2)^2}\Bigg).
\end{align*}
Lines (\ref{I3'}$+$\ref{I4'}) are computed directly to avoid a singularity at $p_1=p_2$, and is given by
\begin{align*}
 &\frac{2}{(1+p_1+p_2)^2(p_1-p_2)}\Bigg(\frac{\log(1+p_2)}{p_2(1+2p_2)(1+p_2)}-\frac{\log(1+p_1)}{p_1(1+2p_1)(1+p_1)}\\
&\qquad\qquad\qquad\qquad\qquad+\frac{(2+3p_1+p_2)\left(\frac{\pi^2}{6}-\log(1+p_1)^2-2\mathrm{Li}_2\left(-p_1\right)\right)}{(1+2p_1)^2(1+p_1+p_2)}\\
&\qquad\qquad\qquad\qquad\qquad-\frac{(2+3p_2+p_1)\left(\frac{\pi^2}{6}-\log(1+p_2)^2-2\mathrm{Li}_2\left(-p_2\right)\right)}{(1+2p_2)^2(1+p_1+p_2)}\Bigg).
\end{align*}
Adding all terms and using the well known identity 
\begin{align}\label{dilogidentity}
 \mathrm{Li}_2\left(-x\right)+\frac{1}{2}\log(1+x)^2=-\mathrm{Li}_2\left(\frac{x}{1+x}\right),
\end{align}
the result is then given by
\begin{align}\label{G4}
	G^{\quad aa}_{4,\,p_1p_2}&=\frac{2}{(1+p_1+p_2)^2(p_1-p_2)}\Bigg(\frac{3\log(\frac{1+p_1}{1+p_2})^2}{(1+p_1+p_2)(p_1-p_2)}\nonumber\\
	&+\frac{2\log(1+p_1)}{p_1(1+2p_1)(1+p_1)}-\frac{2\log(1+p_2)}{p_2(1+2p_2)(1+p_2)}\\
	&-\frac{(1+2p_2)\left(\frac{\pi^2}{6}+2\mathrm{Li}_2(\frac{p_1}{1+p_1})\right)}{(1+2p_1)^2(1+p_1+p_2)}
	+\frac{(1+2p_1)\left(\frac{\pi^2}{6}+2\mathrm{Li}_2(\frac{p_2}{1+p_2})\right)}{(1+2p_2)^2(1+p_1+p_2)}\Bigg).\nonumber
\end{align}
Equation (\ref{G4}) is confirmed by loop expansion in Appendix \ref{appendixA}.

\addsec{\large{Order $n=3$}}
The 2-point functions $G^{\quad aa}_{0,\,p_1p_2},\,G^{\quad aa}_{2,\,p_1p_2}$ and $G^{\quad aa}_{4,\,p_1p_2}$ are inserted into (\ref{rekursiveGL}).
We split the integrals into individual parts, which certainly converge. The identity (\ref{dilogidentity}) is used to achieve terms of the form
$\mathrm{Li}_2\left(-x\right)$, which are easier to integrate. 
With the definition of the trilogarithm 
\begin{align*}
 \mathrm{Li}_3(-x)=\int_{0}^{x}du\frac{\mathrm{Li}_2(-u)}{u},
\end{align*}
and the identities
\begin{align*}
 \mathrm{Li}_3(-x)=&\mathrm{Li}_3\left(-\,\frac{1}{x}\right)-\frac{1}{6}\log(x)^3-\frac{\pi^2}{6}\log(x)\\
 \mathrm{Li}_3(-x)=&-\!\mathrm{Li}_3\!\!\left(\!\frac{x}{1+x}\!\right)\!+\frac{\log(1+x)^3}{3}-\frac{\log(x)\log(1+x)^2}{2}
 -\frac{\pi^2\log(1+x)}{6}+\!\zeta(3),
\end{align*}
where $\zeta(x)$ is the Riemannian $\zeta$ function, we finally find the correlation function of order $\lambda^6$ 
\begin{align}\label{G6}
 &G^{\quad aa}_{6,\,p_1p_2}=\Big\{\log(1+p_1)f_1(p_1,p_2)+\pi^2\log(1+p_1)f_2(p_1,p_2)+\log(1+p_1)^2f_3(p_1,p_2)\nonumber\\
 &+\left(\mathrm{Li}_2(\tfrac{p_1}{1+p_1})+\tfrac{\pi^2}{6}\right)f_4(p_1,p_2)+\text{Li}_2(\tfrac{p_1}{1+p_1}) \log (\tfrac{1+p_1}{1+p_2})f_5(p_1,p_2)\\
 &+ \!\left(\!\text{Li}_3(-p_1)+\text{Li}_3(\tfrac{p_1}{1+p_1})+\text{Li}_2(\tfrac{p_1}{1+p_1}) 
 \log (1+p_1)+\tfrac{\log (1+p_1)^3}{6} -\tfrac{\pi ^2 \log (1+p_1)}{6}\right)\!f_6(p_1,p_2)\nonumber\\
 &+\left(\text{Li}_3(\tfrac{p_1}{1+p_1})+\tfrac{\pi ^2 \log (1+p_1)}{3} \right)f_7(p_1,p_2)\Big\}+\{p_1 \leftrightarrow p_2\}\nonumber\\
 &+\log(\tfrac{1+p_1}{1+p_2})^3f_8(p_1,p_2)+\log(1+p_1)\log(1+p_2)f_9(p_1,p_2)+\pi^2f_{10}(p_1,p_2)\nonumber\\
 &+\pi^2\log(2)f_{11}(p_1,p_2) +\zeta(3)f_{12}(p_1,p_2)\nonumber
\end{align}
with
\begin{align*}
 &f_1(p_1,p_2)= -\frac{8}{p_1(1+p_1)^2(1+2p_1)^2(p_1-p_2)(1+p_1+p_2)^2}\\
 &f_2(p_1,p_2)= \frac{4\left\{(p_1-p_2)^3+(p_1+p_2+1) \left(7 (p_1+p_2+1)^2-3 (2 p_2+1) (p_1-p_2)\right)\right\}}{(1+2p_1)^3(p_1-p_2)(1+p_1+p_2)^4(1+2p_2)^2}\\
 &f_3(p_1,p_2)= \frac{6}
 {p_1^2(1+p_1)^2(1+2p_1)^3(p_1-p_2)^2(1+p_1+p_2)^3}\\
 &\times\!\!\big\{\! (1 + p_1 + 
     p_2) (2 (p_1 - p_2) (1 + 10 p_1 (1 + p_1)) + 3 (1 + p_1) (1 + 2 p_1))\\
     &\qquad+2 p_1 (1 + p_1) (1 + 2 p_1)^2 \big\}\\
 &f_4(p_1,p_2)=-\frac{4}{p_1(1+p_1)^2(1+2p_1)^3(p_1-p_2)^2(1+p_1+p_2)^3}\\
 &\times\big\{ (1 + p_1 + p_2) (2 (1 + p_2) + 
     p_1 (11 + 43 p_1 + 38 p_1^2 - 6 (3 + 4 p_1) p_2))\\
 &\qquad-2 (1 + p_1) (1 + 2 p_1)^3  \big\}\\
 &f_5(p_1,p_2)=\frac{12\left\{ 2 (p_1-p_2)^2+(2 p_2+1) (p_1+p_2+1)\right\}}{(1+2p_1)^2(p_1-p_2)^3(1+p_1+p_2)^4}\\
 &f_6(p_1,p_2)=-\frac{24\left\{ (1 + p_1 + p_2) (10 (p_1 - p_2)^2 + (1 + 3 p_1 - p_2)^2) - (p_1 - p_2)^3\right\}}{(1+2p_1)^4(p_1-p_2)^3(1+p_1+p_2)^3}\\
 &f_7(p_1,p_2)=-\frac{12\left\{ 5+6p_1+4p_2\right\}}{(1+2p_1)^4(p_1-p_2)(1+p_1+p_2)^3}\\
 &f_8(p_1,p_2)=\frac{20}{(1+p_1+p_2)^4(p_1-p_2)^3}\\
 &f_9(p_1,p_2)=-\frac{24\left\{2 p_1^2-2 p_1 p_2+p_1+2 p_2^2+p_2\right\}}{p_1(1+p_1)(1+2p_1)(p_1-p_2)^2p_2(1+p_2)(1+2p_2)(1+p_1+p_2)^2}\\
 &f_{10}(p_1,p_2)=\frac{4}{3p_1(1+p_1)(1+2p_1)^3p_2(1+p_2)(1+2p_2)^3(1+p_1+p_2)^3}\\*
 &\times\big[ p_1 p_2 \big\{(p_1+p_2+1) \big(48 p_1^3+(-48 p_1^2-24 p_1+72) p_2^2+(-40 p_1^2-12 p_1+56) p_2\\*
 &\qquad\qquad+ 88 p_1^2+56 p_1+32 p_2^3+24\big)
 -(2 p_1+1)^2 (4 p_1 (p_1+1)-1)\big\}\\*
 &\qquad\qquad+2 (2 p_1+1)^2 (2 p_2+1)^2 (p_1+p_2+1)^3\\*
 &\qquad\qquad+p_1 (p_1+1) (2 p_1+1)^2+p_2 (p_2+1) (2 p_2+1)^2\big]\\*
 &f_{11}(p_1,p_2)=-\frac{32\left\{9 (p_1-p_2)^2+7 (p_1+p_2+1)^2\right\}}{(1+2p_1)^4(1+2p_2)^4(1+p_1+p_2)}\\
 &f_{12}(p_1,p_2)=\frac{24\left\{(p_1-p_2)^2+5 (p_1+p_2+1)^2\right\}}{(1+2p_1)^3(1+2p_2)^3(1+p_1+p_2)^3},
\end{align*}
where $\{p_1\leftrightarrow p_2\}$ are the first seven terms by
interchanging $p_1$ and $p_2$. To obtain these results we 
computed a primitive of the integrals using a computer algebra system 
and took the limits $q\to 0$ and $q \to \infty$ by hand. More than 
20 different type of loops contribute at sixth order in $\lambda$, 
and (\ref{G6}) is the sum of all of them.

Most of the terms in (\ref{G6}) are individually divergent 
in the limit $p_2\to p_1$ or $p_{1/2}\to 0$. However, in both limits $G^{\quad aa}_{6,\,p_1p_2}$ has a finite result. For
$p=p_1=p_2$ we find
\begin{align}\label{Gp6}
 G^{\quad\!\! aa}_{6,\,pp}&=\frac{1776}{(1+2p)^7}\big\{\text{Li}_3(-p)+\frac{93}{74} \text{Li}_3(\tfrac{p}{p+1})+
 \text{Li}_2(\tfrac{p}{p+1}) \log (p+1)\nonumber\\
 &\qquad\qquad\qquad +\frac{1}{6} \log ^3(p+1)-\frac{14}{111} \pi ^2 \log (p+1)-\frac{14}{111}  \pi ^2 \log (2)+\frac{5}{74} \zeta (3)\big\}\nonumber\\
 &+
 \frac{2\pi^2(10 p (p (4 p+39)+60)+257)}{3(1+p)^3(1+2p)^6}+\frac{2(9+10p)}{p(1+2p)^7}+\frac{4\log(1+p)(5+7p)}{p^2(1+p)^2(1+2p)^5}\\
 &-\frac{2\log(1+p)^2(p (p+1) (546 p (p+1)+125)+11)}{p^3(1+p)^3(1+2p)^6}\nonumber\\
 &+\frac{4\mathrm{Li}_2(\tfrac{p}{1+p})(7+(1+p)(176 p^3+75 p^2-44 p-11))}{p^2(1+p)^3(1+2p)^6}.\nonumber
\end{align}
It is nice to see how in (\ref{Gp6}) the linear divergence for $p\to 0$ in the last
four terms cancels perfectly, since 
$\lim_{p\to 0}\frac{\mathrm{Li}_2(\tfrac{p}{1+p})}{p^2}=\frac{1}{p}+\mathcal{O}(1)$. Also remarkable to assess is the fact that all kind of functions 
appearing the first time at order $\lambda^6$ (first two lines of (\ref{Gp6})) have the same dependence of $p$ in the denominator.

Sending $p_2\to p_1$ in (\ref{GeschlosseneGL}) an integral equation with derivatives $\frac{\partial G^{aa}_{p_1q}}{\partial p_1}$
appears. Making use of all results (\ref{Loop2}),(\ref{G4}) and (\ref{G6}), the numerical solution for the 2-point function with 
zero momenta can be given up to the eighth-order 
\begin{align*}
 G^{aa}_{00}= &1+2\lambda^2+2(\pi^2-6)\lambda^4+\left\{\pi^2\left(\tfrac{514}{3}-224 \log(2)\right)+120\,\zeta(3)-266\right\} \lambda^6\\*
 &+194.612 \,\lambda^8+\mathcal{O}(\lambda^{10}).
\end{align*}

\section{Conclusion and outlook}

We have introduced the noncommutative 3-colour model as a
quantum field theoretical model in two dimensions. 
We derived the Schwinger-Dyson equations of the 2-point function 
and the $N$-point functions for a single boundary component.
This required a generalisation of the Ward-Takahashi identity to 
coloured models. This new identity 
seems to be related to a mixing symmetry between two colours.  
In the large $\mathcal{N},V$ limit a
closed integral equation (\ref{GeschlosseneGL}) occurs, which is a
\textit{non-perturbative} result.  This equation was used to find
perturbative solutions up to the sixth order in the coupling constant.

The main aim for the future is
to find an exact solution of (\ref{GeschlosseneGL}) or to prove
existence, if possible also uniqueness, of a
solution. Furthermore, we want to extend this work to
determine Schwinger-Dyson equations for $B\geq 2$, where already
problems arise for the $(1+1)$-point function. Finally, we would like
to treat the renormalisation problems in dimension 4 and 6.

\appendix
\section{$G^{\quad\! aa}_{4,\,p_1p_2}$ computation by graphs}\label{appendixA}
 Representatives of all graphs with one boundary component and two external edges at the fourth order in $\lambda$ are the following 
 graphs:\vspace*{2ex} \\
\vspace*{4ex}
\raisebox{4ex}{$\Gamma_1:$}\qquad\quad
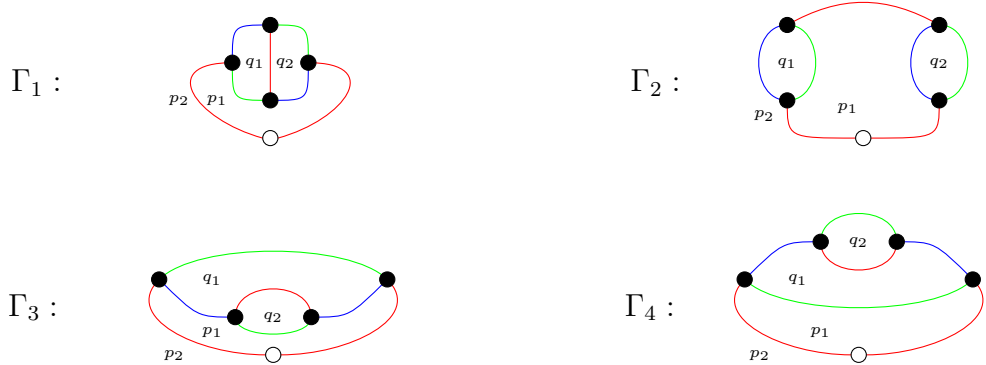
\begin{tikzpicture}
\draw [draw=red] (2.9,0) .. controls (2,0.25) and (1.5,1).. (2.5,1) node at (2.3,0.5) {\tiny $p_1$};
\draw [draw=red] (3.1,0) .. controls (4,0.25) and (4.5,1).. (3.5,1)  node at (1.8,0.5) {\tiny $p_2$};
\draw [draw=blue] (3,1.5) .. controls (2.5,1.5) and (2.5,1.5).. (2.5,1) node at (2.8,1) {\tiny $q_1$};
\draw [draw=green] (3,1.5) .. controls (3.5,1.5) and (3.5,1.5).. (3.5,1) node at (3.2,1) {\tiny $q_2$};
\draw [draw=green] (3,0.5) .. controls (2.5,0.5) and (2.5,0.5).. (2.5,1);
\draw [draw=blue] (3,0.5) .. controls (3.5,0.5) and (3.5,0.5).. (3.5,1);
\draw [draw=red] (3,0.5) -- (3,1.5);
\filldraw (3.5,1) circle (0.1cm);
\filldraw (2.5,1) circle (0.1cm);
\filldraw (3,1.5) circle (0.1cm);
\filldraw (3,0.5) circle (0.1cm);
\draw (3,0) circle (0.1cm);
\end{tikzpicture}
\qquad\qquad\qquad\quad\,
\raisebox{4ex}{$\Gamma_2:$}\qquad
\begin{tikzpicture}
\draw [draw=red] (2.9,0) .. controls (2,0) and (2,0).. (2,0.5) node at (2.8,0.4) {\tiny $p_1$};
\draw [draw=red] (3.1,0) .. controls (4,0) and (4,0).. (4,0.5) node at (1.7,0.3) {\tiny $p_2$};
\draw [draw=blue] (2,0.5) .. controls (1.5,0.5) and (1.5,1.5).. (2,1.5) node at (2.,1) {\tiny $q_1$};
\draw [draw=green] (4,0.5) .. controls (4.5,0.5) and (4.5,1.5).. (4,1.5) node at (4.,1) {\tiny $q_2$};
\draw [draw=blue] (4,0.5) .. controls (3.5,0.5) and (3.5,1.5).. (4,1.5);
\draw [draw=green] (2,0.5) .. controls (2.5,0.5) and (2.5,1.5).. (2,1.5);
\draw [draw=red] (2,1.5) .. controls (2.7,1.9) and (3.3,1.9).. (4,1.5);
\filldraw (2,0.5) circle (0.1cm);
\filldraw (4,0.5) circle (0.1cm);
\filldraw (2,1.5) circle (0.1cm);
\filldraw (4,1.5) circle (0.1cm);
\draw (3,0) circle (0.1cm);
\end{tikzpicture}\\
\vspace*{4ex}
\raisebox{4ex}{$\Gamma_3:$}\qquad
\begin{tikzpicture}
\draw [draw=red] (2.9,0) .. controls (2,0) and (1,0.5).. (1.5,1) node at (2.2,0.3) {\tiny $p_1$};
\draw [draw=red] (3.1,0) .. controls (4,0) and (5,0.5).. (4.5,1) node at (1.7,0.) {\tiny $p_2$};
\draw [draw=green] (1.5,1) .. controls (2,1.5) and (4,1.5).. (4.5,1) node at (2.2,1) {\tiny $q_1$};
\draw [draw=blue] (1.5,1) .. controls (2,0.5) and (2,0.5).. (2.5,0.5) node at (3.,0.5) {\tiny $q_2$};
\draw [draw=blue] (4.5,1) .. controls (4,0.5) and (4,0.5).. (3.5,0.5);
\draw [draw=red] (2.5,0.5) .. controls (2.5,1) and (3.5,1).. (3.5,0.5);
\draw [draw=green] (2.5,0.5) .. controls (2.5,0.2) and (3.5,0.2).. (3.5,0.5);
\filldraw (1.5,1) circle (0.1cm);
\filldraw (4.5,1) circle (0.1cm);
\filldraw (3.5,0.5) circle (0.1cm);
\filldraw (2.5,0.5) circle (0.1cm);
\draw (3,0) circle (0.1cm);
\end{tikzpicture}
\qquad\qquad\qquad
\raisebox{4ex}{$\Gamma_4:$}\quad
\begin{tikzpicture}
\draw [draw=red] (2.9,0) .. controls (2,0) and (1,0.5).. (1.5,1) node at (2.5,0.3) {\tiny $p_1$};
\draw [draw=red] (3.1,0) .. controls (4,0) and (5,0.5).. (4.5,1) node at (1.7,0.) {\tiny $p_2$};
\draw [draw=green] (1.5,1) .. controls (2,0.5) and (4,0.5).. (4.5,1) node at (2.2,1) {\tiny $q_1$};
\draw [draw=blue] (1.5,1) .. controls (2,1.5) and (2,1.5).. (2.5,1.5) node at (3.,1.5) {\tiny $q_2$};
\draw [draw=blue] (4.5,1) .. controls (4,1.5) and (4,1.5).. (3.5,1.5);
\draw [draw=green] (2.5,1.5) .. controls (2.5,2) and (3.5,2).. (3.5,1.5);
\draw [draw=red] (2.5,1.5) .. controls (2.5,1) and (3.5,1).. (3.5,1.5);
\filldraw (1.5,1) circle (0.1cm);
\filldraw (4.5,1) circle (0.1cm);
\filldraw (3.5,1.5) circle (0.1cm);
\filldraw (2.5,1.5) circle (0.1cm);
\draw (3,0) circle (0.1cm);
\end{tikzpicture}.\\
Let $a= a_1^1=a_2^1$.
With straightforward computation by using the introduced rules in Section \ref{SectionGraph} one finds
\begin{align*}
	\tilde{G}^{aa}_{p_1p_2}&(\Gamma_1)=\frac{\lambda^4}{(1+p_1+p_2)^2}\int_{0}^{\infty}\!\!\!
	\frac{\frac{dq_1dq_2}{1+q_1+q_2}}{(1+p_1+q_1)(1+p_1+q_2)(1+p_2+q_1)(1+p_2+q_2)}\nonumber\\
	=&\frac{\lambda^4}{(1+p_1+p_2)^2}\Bigg(-\frac{\log(1+p_1)^2}{(p_1-p_2)^2(1+2p_1)}-\frac{\log(1+p_2)^2}{(p_1-p_2)^2(1+2p_2)}\nonumber\\
	&-\frac{\pi^2/6-2\mathrm{Li}_2\left(-p_1\right)}{(1+2p_1)(p_1-p_2)(1+p_1+p_2)} 
	+\frac{\pi^2/6-2\mathrm{Li}_2\left(-p_2\right)}{(1+2p_2)(p_1-p_2)(1+p_1+p_2)}\nonumber\\
	&+\frac{2\log(1+p_1)\log(1+p_2)}{(p_1-p_2)^2(1+p_1+p_2)}\Bigg)\nonumber\\
	\tilde{G}^{aa}_{p_1p_2}&(\Gamma_2)=\frac{\lambda^4}{(1+p_1+p_2)^3}\int_{0}^{\infty}\!\!\!
	\frac{dq_1dq_2}{(1+p_1+q_1)(1+p_2+q_1)(1+p_2+q_1)(1+p_2+q_2)}\nonumber\\
	=&\frac{\lambda^4}{(1+p_1+p_2)^3}\frac{\log(\frac{1+p_1}{1+p_2})^2}{(p_1-p_2)^2}\nonumber\\
	\tilde{G}^{aa}_{p_1p_2}&(\Gamma_3)=\frac{\lambda^4}{(1+p_1+p_2)^2}\int_{0}^{\infty}\frac{\frac{dq_1dq_2}{1+q_1+q_2}}{(1+p_1+q_1)^2(1+p_1+q_2)(1+p_2+q_1)}\nonumber\\*
	=&\frac{\lambda^4}{(1+p_1+p_2)^2}\Bigg(-\frac{  \mathrm{Li}_2\left(-p_1\right)  }{(1+2p_1)^2(1+p_1+p_2)}
	-\frac{\frac{\pi^2}{6}-\log(1+p_1)^2- \mathrm{Li}_2\left(-p_1\right) }{(1+2p_1)(p_1-p_2)^2}\nonumber\\*
	&-\frac{\frac{\pi^2}{6}-\log(1+p_1)^2- \mathrm{Li}_2\left(-p_1\right)  }{(1+2p_1)^2(p_1-p_2)}
	+\frac{\frac{\pi^2}{6}-\log(1+p_2)\log(1+p_1)- \mathrm{Li}_2\left(-p_2\right) }{(1+p_1+p_2)(p_1-p_2)^2}\nonumber\\*
	&+\frac{\log(1+p_1)}{p_1(1+p_1)(1+2p_1)(p_1-p_2)}\Bigg)\nonumber\\
	\tilde{G}^{aa}_{p_1p_2}&(\Gamma_4)=\frac{\lambda^4}{(1+p_1+p_2)^2}\int_{0}^{\infty}\frac{\frac{dq_1dq_2}{1+q_1+q_2}}{(1+p_1+q_1)(1+p_2+q_1)^2(1+p_2+q_2)}\nonumber\\
	=&\frac{\lambda^4}{(1+p_1+p_2)^2}\Bigg(-\frac{  \mathrm{Li}_2\left(-p_2\right)  }{(1+2p_2)^2(1+p_1+p_2)}
	-\frac{\frac{\pi^2}{6}-\log(1+p_2)^2- \mathrm{Li}_2\left(-p_2\right) }{(1+2p_2)(p_1-p_2)^2}\nonumber\\
	&+\frac{\frac{\pi^2}{6}-\log(1+p_2)^2- \mathrm{Li}_2\left(-p_2\right)  }{(1+2p_2)^2(p_1-p_2)}
	+\frac{\frac{\pi^2}{6}-\log(1+p_2)\log(1+p_1)- \mathrm{Li}_2\left(-p_1\right) }{(1+p_1+p_2)(p_1-p_2)^2}\nonumber\\
	&-\frac{\log(1+p_2)}{p_2(1+p_2)(1+2p_2)(p_1-p_2)}\Bigg).\nonumber
\end{align*}

We verify easily that $s(\Gamma_1)=2,\,s(\Gamma_2)=4,\, s(\Gamma_3)=4$ and $ s(\Gamma_4)=4$. The correlation function of order $\lambda^4$
is finally given with identity (\ref{dilogidentity}) by
\begin{align*}
	G^{\quad aa}_{4,\,p_1p_2}&=\sum_{i=1}^4 s(\Gamma_i)\tilde{G}^{aa}_{p_1p_2}(\Gamma_i)\\
	&=\frac{2}{(1+p_1+p_2)^2(p_1-p_2)}\Bigg(\frac{3\log(\frac{1+p_1}{1+p_2})^2}{(1+p_1+p_2)(p_1-p_2)}\nonumber\\
	&+\frac{2\log(1+p_1)}{p_1(1+2p_1)(1+p_1)}-\frac{2\log(1+p_2)}{p_2(1+2p_2)(1+p_2)}\\
	&-\frac{(1+2p_2)\left(\frac{\pi^2}{6}+2\mathrm{Li}_2(\frac{p_1}{1+p_1})\right)}{(1+2p_1)^2(1+p_1+p_2)}
	+\frac{(1+2p_1)\left(\frac{\pi^2}{6}+2\mathrm{Li}_2(\frac{p_2}{1+p_2})\right)}{(1+2p_2)^2(1+p_1+p_2)}\Bigg).\nonumber
\end{align*}

\subsection*{Acknowledgement}
This work was supported by the Deutsche Forschungsgemeinschaft (SFB 878). 
A. H. wants to thank Jins de Jong for helpful discussions.

\addcontentsline{toc}{section}{References}
\bibliographystyle{ieeetr}
\bibliography{./Bibo1}

\end{document}

%% file: Pakete.tex
\usepackage[utf8]{inputenc}
\usepackage{ngerman}
\usepackage{textcomp}

\usepackage{epsfig}
\usepackage{graphicx}
\usepackage{floatflt}
\usepackage{float} 		
\usepackage{sidecap}
\usepackage{wrapfig}
\usepackage{tikz}

\usepackage{dsfont}
\usepackage{amssymb}
\usepackage{amsthm}
\usepackage{amsmath}
\usepackage{nicefrac}
\usepackage{enumitem}
\usepackage{mathtools}

\newtheorem{lemma}{Lemma}

\newtheorem{proposition}{Proposition}

\usepackage{hyperref}
\usepackage{geometry}
\usepackage[numbers]{natbib}

\usepackage[textwidth=3cm, 			
			textsize=footnotesize, 	
			shadow, 				
			colorinlistoftodos 		
			]{todonotes}

\usepackage[german, english]{babel}		
			